\documentstyle[12pt]{article}
\input html.sty
\begin{document}
\title{MATTERS OF GRAVITY, The newsletter of the APS Topical Group 
on Gravitation}
\begin{center}
{ \Large {\bf MATTERS OF GRAVITY}}\\
\bigskip
\hrule
\medskip
{The newsletter of the Topical Group on Gravitation of the American Physical 
Society}\\
\medskip
{\bf Number 12 \hfill Fall 1998}
\end{center}
\begin{flushleft}

\tableofcontents
\bigskip
\hrule
\vfill
\section*{\noindent Editorial policy:}

The newsletter publishes articles in three broad categories, 

1. News about the topical group, normally contributed by officers of the 
group.

2. Research briefs, comments about new developments in research,
typically by an impartial observer. These articles are normally
by invitation, but suggestions for potential topics and authors are welcome
by the correspondents and the editor.

3. Conference reports, organizers are welcome to contact the editor or
correspondents, the reports are sometimes written by participants in the
conference in consultation with organizers.

Articles are expected to be less than two pages in length in all categories.

Matters of Gravity is not a peer-reviewed journal for the publication
of original research. We also do not publish full conference or meeting
announcements, although  we might consider publishing a brief notice 
with indication of a web page or other contact information.

\section*{\noindent  Editor\hfill}
%{\bf Editor:}

\medskip
Jorge Pullin\\
\smallskip
Center for Gravitational Physics and Geometry\\
The Pennsylvania State University\\
University Park, PA 16802-6300\\
Fax: (814)863-9608\\
Phone (814)863-9597\\
Internet: 
\htmladdnormallink{\protect {\tt pullin@phys.psu.edu}}
{mailto:pullin@phys.psu.edu}\\
WWW: \htmladdnormallink{\protect {\tt http://www.phys.psu.edu/\~{}pullin}}
{http://www.phys.psu.edu/~pullin}
\begin{rawhtml}
<P>
<BR><HR><P>
\end{rawhtml}
%{\bf \Large Contents:}
\end{flushleft}
\pagebreak
\section*{Editorial}

I wanted to mention two topics that I find exciting. First, as mentioned in
Jim Isenberg's article, the TGG (notice that this is now the official
acronym) membership is growing. Becoming a division of APS is starting
to appear as less of a dream. I was planning to start a recruitment
drive using the newsletter, but at the moment it seems premature. It
probably is wise to wait to grow a bit more and then try to go all
out to reach that magical 1200 number.

The other topic I wanted to mention is the increasing use of the
internet to transmit voice. Concretely, people are starting to record
and post on the web conferences and lectures, together with the slides
used by the speaker. New audio tools and compression techniques make
this feasible even with not too fast network connections.  As a token
example, the introductory physics lectures I teach at Penn State are
on the web. The ITP at Santa Barbara has the whole Strings conference
and a bunch of other things online, you can find them at
\htmladdnormallink{\protect {\tt http://www.itp.ucsb.edu/online}}
{http://www.itp.ucsb.edu/online}. The Center for Gravitational Physics
and Geometry at PSU is posting its semi-weekly talks online, you can
listen to them at \htmladdnormallink
{\protect {\tt http://vishnu.nirvana.phys.psu.edu/relativity\_seminars.html}}
{http://vishnu.nirvana.phys.psu.edu/relativity\_seminars.html}.  The
importance for research of this tool cannot be understressed. One can
keek up with the forefront of research without leaving one's office.
In this time of one-conference-every-week, this can be a life saver
for physicists who have become road-warriors or for those with limited
travel budgets. I will start a link in the TGG's web page to such
audio gravity resources, if you know of more, please let me know.

The next newsletter is due February 1st.  If everything goes well this
newsletter should be available in the gr-qc Los Alamos archives under
number gr-qc/9809031. To retrieve it send email to 
\htmladdnormallink{gr-qc@xxx.lanl.gov}{mailto:gr-qc@xxx.lanl.gov}
(or 
\htmladdnormallink{gr-qc@babbage.sissa.it}{mailto:gr-qc@babbage.sissa.it} 
in Europe) with Subject: get 9809031
(numbers 2-10 are also available in gr-qc). All issues are available in the
WWW:\\\htmladdnormallink{\protect {\tt
http://vishnu.nirvana.phys.psu.edu/mog.html}}
{http://vishnu.nirvana.phys.psu.edu/mog.html}\\ 
A hardcopy of the newsletter is
distributed free of charge to some members of the APS
Topical Group on Gravitation. It is considered a lack of etiquette to
ask me to mail you hard copies of the newsletter unless you have
exhausted all your resources to get your copy otherwise.

If you have comments/questions/complaints about the newsletter email
me. Have fun.
\bigbreak

\hfill Jorge Pullin\vspace{-0.8cm}
\vfill
\pagebreak

\section*{Correspondents}
\begin{itemize}
\item John Friedman and Kip Thorne: Relativistic Astrophysics,
\item Raymond Laflamme: Quantum Cosmology and Related Topics
\item Gary Horowitz: Interface with Mathematical High Energy Physics and
String Theory
\item Richard Isaacson: News from NSF
\item Richard Matzner: Numerical Relativity
\item Abhay Ashtekar and Ted Newman: Mathematical Relativity
\item Bernie Schutz: News From Europe
\item Lee Smolin: Quantum Gravity
\item Cliff Will: Confrontation of Theory with Experiment
\item Peter Bender: Space Experiments
\item Riley Newman: Laboratory Experiments
\item Warren Johnson: Resonant Mass Gravitational Wave Detectors
\item Stan Whitcomb: LIGO Project
\item Peter Saulson; former editor, correspondent at large.
\end{itemize}
\vfill
\pagebreak

%%%%%%%%%%%%%%%%%%%%%%%%%%%%%%%%
%%%%%%%%%%%%%%%%%%%%%%%%%%%%%%%%
%%%%%%%%%%%%%%%%%%%%%%%%%%%%%%%%

\section*{\centerline {Topical group news}}
\addtocontents{toc}{\protect\smallskip}
\addtocontents{toc}{\bf News:}
\addtocontents{toc}{\protect\smallskip}
\addcontentsline{toc}{subsubsection}{\it  Topical group news, by Jim Isenberg}
\begin{center}
    Jim Isenberg, TGG secretary, University of Oregon\\
\htmladdnormallink{jim@newton.uoregon.edu}
{mailto:jim@newton.uoregon.edu}
\end{center}
\parindent=0pt
\parskip=5pt

Since the summer is a relatively quiet time for TGG activity, not  
much has happened since the April meeting in Columbus. So most of our  
news is contained in the following excerpts from the minutes of the  
Columbus meeting.

1) Officers and Committees.

The  list of officers and committee members for the coming year are as  
follows:

Chair: Abhay Ashtekar

Chair Elect: Rainer Weiss

Vice Chair: Cliff Will

Secretary/Treasurer (1996-1999): Jim Isenberg

Delegates (1998-2001): Steve Carlip, Peter Saulson

Delegates (1997-2000): Lee Samuel Finn, Mac Keiser

Delegates (1996-1999): Frederick Raab, Leonard Parker

Nominating Committee: not yet chosen

Fellowship Committee: Cliff Will,  Bill Hamilton, Richard Price

Program Committee: Rai Weiss, David Shoemaker, Beverly Berger

Editor MOG and Webmaster: Jorge Pullin

2) On New Members.

The TGG is a very rapidly growing, at least so far. We have about 530  
members. It was noted that if we get to 1200 (which is approximately  
3\% of the total APS membership) we could qualify as a division. One  
source of new members we might push is our students.The first year of  
a student membership is free.

3) Hartle Committee Discussion.

Every ten years, the National Research Council sponsors an in-depth  
assessment of the past accomplishments, current state, and future  
prospects of physics research in a number of fields. This year, for  
the first time, a committee was appointed to prepare a separate book  
on gravitational physics.. Jim Hartle, who chairs  this committee,  
led a discussion at the general meeting, with the emphasis on  
soliciting opinions from the members of the TGG regarding the state  
of gravitational physics research. There was discussion on the  
adequacy of research support; on interdisciplinary work with data  
analysis people, math people, computer people, and optics people; on  
the possibility of instituting summer schools focussing on  
gravitational physics (like TASI for the particle physics people, and  
the AMS-run summer schools in math), and a number of other topics.  
Jim Hartle noted that the deadline for input is essentially the  
beginning of May, and then the report will be put together.

4) Centenary Meeting.

The next APS ``April Meeting" will be in March
1999 in Atlanta, and it will mark the APS' Centenary. This will be the
TGG's annual meeting, and plans for the meeting were discussed. We are
promised 2 invited sessions, plus one ``Centennial Symposium". The
``Centennial Symposia" are essentially the same as the focus sessions
of this year's meeting, but intended for a broader audience. Beverly
Berger will be our liaison with APS in scheduling the Centennial
Symposia.  Rai Weiss will be our liaison for the general program,
including the invited sessions.

There was a report by Phil Lindquist on the proposed display for the
Centennial meeting. As approved earlier by members of the Executive
Committee, it will focus on gravitational radiation --sources,
detection, history of ideas-- with LIGO featured. We have been
allotted an 8 x 10 space, although we may argue for a bigger space. A
rough tentative plan was shown. The tentative cost --\$10,000-- was
discussed. After much discussion, the committee generally agreed to
support one third of the cost, up to \$3000. LIGO will pay for the
rest.  Some people noted that the rough sketch of the display
contained too much material. All agreed that it will be important to
make the display accessible, both to those interested only in a quick
perusal, and those who want to learn about gravitational radiation in
a bit more detail.

5) Viewpoint on Our Role in APS and These Meetings.

Since the APS meetings are expensive, and consequently not that well  
attended,  the possibility was brought up that we might have the TGG  
annual meeting at one of the 3 regional meetings on a rotating basis.  
This issue brought up the question of what the role of the TGG is,  
regarding the gravitational researcher community, and the rest of the  
physics community. All at the meeting supported the idea that a major  
function of this group is to raise the visibility of gravitational  
physics. This led to strong support for keeping our meeting at one of  
the large APS meetings. It also led to the view that we have some  
stake in the continued existence of the April meeting, and that we  
should be involved in its evolution.

This led to the issue of attracting more people, especially students,
to the April APS meetings. There was general agreement that we should
spend some of our surplus to provide travel support to students. This
could start next year, although no plan for implementing the
suggestion was discussed.

6) Prizes

Since last year, some of the members of the Executive
Committee have been discussing and researching the possibility of
instituting an APS sponsored prize for gravitational physics. Abhay
Ashtekar outlined the various possibilities allowed by the APS:

i) Senior Prize: These are tightly controlled by the APS. They want
them to be around \$10,000, to be awarded at least every other year,
and they want very strong assurances that we have 10 years of worthy
candidates. So to do this sort of prize would require a lot of paper
work, and a lot of fund raising (at least \$200,000)
ii) Junior Prize: This could be an early career or post-doc
award. Less funding and justification is needed.
iii) Dissertation Prize: These are not strongly controlled. We could
likely institute one without too much work or fund-raising.

While the senior award would take by far the most effort to institute,
there was strong support for it. The justification was that it would
be the best way to enhance the visibility of the field. Since funding
is a big issue, it was decided to pursue discrete, quiet, attempts to
locate possible funding sources. Such sources need to be identified
before proceeding further with plans to set up a prize.

It was noted that we might also try to set up a junior level prize,  
but wait on this until the funding inquiries have been made.

\vfill
\pagebreak
%%%%%%%%%%%%%%%%%%%%%%%%%%%%%%%%%%%%%%%%%%
\section*{\centerline {Summer school in gravitational physics opportunity}}
\addcontentsline{toc}{subsubsection}{\it  
Summer school in gravitational physics opportunity, by Jim Hartle}
\begin{center}
    Jim Hartle, UC Santa Barbara\\
\htmladdnormallink{hartle@big-g.physics.ucsb.edu}
{hartle@big-g.physics.ucsb.edu}
\end{center}
\parindent=0pt
\parskip=5pt

In taking input for the Committee on Gravitational Physics one 
theme which emerged a number of times was the utility of an
an occasional or regular Gravitational Physics Summer School.

\noindent Some positive arguments for such a school were:
\begin{itemize}
\item It would help unify the field.
\item It would provide an advanced pedagogical literature.
\item In an subject that is becoming increasingly  integrated with other
     parts of physics, it could broaden students and postdocs 
     in the area by acquainting them with developments related areas
     such as astrophysics and particle physics.   
\item It would provide a way of introducing the field to students in related
     areas at universities where there are not  strong relativity
     groups, and perhaps recruiting some talent into the field. 
\end{itemize}

\noindent Some of the negative comments were:
\begin{itemize}
\item It is not really needed, because, unlike, say particle
  physics, students do not really need much advanced training to start
  work on their Ph.D.
\item  The occasional European schools, e.g. Les Houches 1992, are enough.
\end{itemize}

The ITP at Santa Barbara is undertaking to help ``nucleate''
summer schools. What they mean by that is that if suitable outside
funding were found, they might run a school once or perhaps irregularly
as a kind of demonstration project.  They would do the basic local
work of arranging accommodation, handling applications, etc which 
they are set up to do. I spoke to the current director
David Gross, and he was receptive, but not of course committal.  
The next available slot is July 2000, and a competition between 
proposals will be held at the advisory board meeting February 1999.

However, the ITP does not itself have the funds to run such a school
which are of order \$50k. That would have to be raised from say
the NSF. Rich Isaacson is supportive but not of course committal.
He would entertain proposals for the necessary funding, either
at the ITP or elsewhere.   

What would be needed to take advantage of the opportunity at the ITP
is a proposal to them say by November of this year with some understanding
from an agency about funding it.  

I'd be pleased to discuss this further if it is of interest.  

\vfill
\pagebreak
%%%%%%%%%%%%%%%%%%%%%%%%%%%%%%%%%%%%%%%%%%
\section*{\centerline {
Bogart, Bergman, and (Al)bert:  What might have been}}
\addcontentsline{toc}{subsubsection}{{\it  
Bogart, Bergman, and (Al)bert, by Cliff Will
and Robert Riemer}} 
\begin{center}
    Clifford Will, Washington University, St. Louis, Robert Riemer$^1$, NRC\\
\htmladdnormallink{cmw@howdy.wustl.edu}
{cmw@howdy.wustl.edu}
\end{center}
\parindent=0pt
\parskip=5pt

The recent list of the top 100 American films put out by the American Film
Institute places {\it Casablanca} in the top 10, a ranking few would
dispute.  Who can forget the haunting song which Ingrid Bergman asked
to be ``played again'', and which brought back such painful memories for
Bogart's Rick:  ``You must remember this, a kiss is still a kiss,
$\dots$''?  

However, few realize that what was sung in the movie is only part
of a song written by Herman Hupfield twelve years earlier.  The intro
to the song goes as follows$^2$:

\medskip
{\parskip=0pt
\begin{itemize}
\item[] This day and age we're living in
\item[] Give cause for apprehension
\item[] With speed and new invention
\item[] And things like fourth dimension
\item[]
\item[] Yet we get a trifle weary
\item[] With Mr. Einstein's theory
\item[] So we must get down to earth at times
\item[] Relax, relieve the tension
\item[]
\item[] And no matter what the progress
\item[] Or what may yet be proved
\item[] The simple facts of life are such
\item[] They cannot be removed
\item[] 
\item[] You must remember this $\dots$
\end{itemize}
}

\medskip
\noindent
What a missed opportunity to connect Albert to Bogie and Bergman!

\bigskip
\noindent
$^1$ Robert (Roc) Riemer is currently NRC staff liaison for the
Committee on Gravitational Physics.

\noindent
$^2$ \copyright 1931 by Warner Bros. Music Corporation, ASCAP

\vfill
\pagebreak
%%%%%%%%%%%%%%%%%%%%%%%%%%%%%%%%%%%%%%%%%%

\section*{\centerline {
New data-analysis subgroups} \\
\centerline{of LIGO Science Collaboration}}
%\addtocontents{toc}{\protect\smallskip}
\addcontentsline{toc}{subsubsection}{\it 
New data-analysis subgroups of LIGO Science Collaboration,
by \'Eanna Flanagan}
\begin{center}
    \'Eanna Flanagan, Cornell \\
\htmladdnormallink{eef3@cornell.edu}
{mailto:eef3@cornell.edu}
\end{center}
\parindent=0pt
\parskip=5pt

With the installation of the initial LIGO interferometers only a
short year or two from now, the LIGO Science Collaboration (LSC) has
started giving more attention to preparing for data analysis.  At the
LSC meeting in March at the LIGO Hanford site, three new subgroups of
the LSC were formed: a {\it Detector Characterization} subgroup, 
a {\it Astrophysical Source Identification and
Signatures} subgroup, and a {\it Detection Confidence and
Statistical Analysis} subgroup.  These subgroups complement the
three previously existing subgroups whose purpose is the design and
development of the detectors themselves (the {\it Stochastic Forces -
Isolation and Suspensions}, {\it Laser and Optics}, and {\it
Interferometer Configurations subgroups}). 

Identifying with high confidence a gravitational wave signal in the
outputs of the LIGO interferometers will involve several elements: (i) A
detailed understanding of the statistical properties of the noise in
the instruments, (ii) computationally efficient algorithms to sift
through the vast amounts of data generated by the instruments to
identify possible candidate signals, with different algorithms for
each type of source, and (iii) for each candidate
signal, exhaustive post-processing analyses of all available
information must be performed in order to assess the probability of it
being a true signal.  The purpose of the three new
subgroups is, roughly speaking, to address these three different
tasks. 

The Astrophysical Source Identification and Signatures subgroup
addresses task (ii) and is chaired by Bruce Allen
(ballen@dirac.phys.uwm.edu).  More information about this subgroup can
be found at its web home-page   

\htmladdnormallink{\protect {\tt
http://www.tapir.caltech.edu/\~{}lsc\_asis/}}
{http://www.tapir.caltech.edu/~lsc_asis/}

which contains links to meetings, research plans etc.  The Detection
Confidence and Statistical Analysis subgroup addresses task (iii) and is
chaired by Sam Finn (finn@phys.psu.edu).  Finally task (i) is the job
of the Detector Characterization subgroup, chaired by Bill Hamilton
(hamilton@phgrav.phys.lsu.edu).  Individuals who are interested in
participating in these groups are encouraged to contact the respective
chairmen.  

\vfill
\pagebreak
%%%%%%%%%%%%%%%%%%%%%%%%%%%%%%%%%%%%%%%%%%
\section*{\centerline {
Marcel Bardon, a man of vision}}
\addcontentsline{toc}{subsubsection}{{\it  
Marcel Bardon, a man of vision, by Richard Isaacson}}
\begin{center}
    Richard Isaacson, National Science Foundation\\
\htmladdnormallink{ri@einstein.mps.nsf.gov}
{ri@einstein.mps.nsf.gov}
\end{center}
\parindent=0pt
\parskip=4pt
Marcel Bardon, Director of the Physics Division of the National Science
Foundation and noted photographer, died on May 20, 1998 at age 72,
after a courageous battle against lymphoma.

Marcel worked at the NSF since 1970, where he served as the Director of
the Physics Division with a few departures to work on international
science. His superb scientific judgment, political skill, and
extraordinary vision allowed him to substantially advanced the broad
frontiers of science. His support for excellent and innovative
projects, often in the face of vocal opposition, has had major impact
upon many fields.  Some of these of particular interest to research in
gravity include his creation of the unique Gravitational Physics
Program at NSF, the building of the Institute for Theoretical
Physics at Santa Barbara, the establishment of the first generation of
NSF Supercomputer Centers, and the construction of the Laser
Interferometer Gravitational-Wave Observatory (LIGO).

Because of his sensitivity to its future potential, the emerging field
of gravitational physics blossomed beyond its traditional theoretical
roots. Marcel began NSF support for cryogenic bar detectors when he
first came to the Foundation as Program Director for Intermediate and
high Energy Physics. With his encouragement as Division Director,
expansion of the new Gravitational Physics Program enabled initial R\&D
on laser interferometers to begin, eventually demonstrating in the
laboratory that a large-scale detector was possible.  Marcel vigorously
championed efforts within the Foundation to proceed with this new
technology. His support ultimately lead to approval for the
construction and operation of LIGO, the largest scientific project the
NSF has undertaken.

Marcel's scientific vision extended well beyond physics, and he
possessed a deep understanding of the fundamental nature of science as
an international endeavor.  During 1979-1981, he served as the
Scientific and Technical Affairs Officer with the UNESCO Mission in
Paris. From 1986-1988, he was NATO's Deputy Assistant Secretary General
for Science in Brussels.  He left the NSF Physics Division for an
extended tour as the Director of the Division of International Programs
during 1992-1997, where he reoriented the division's goals. This lead
to its present emphasis on the initiation of new international
collaborations and encouragement of the early participation of U.S.
scientists in international research.

Marcel's many honors included the NSF Meritorious and Distinguished
Service Medals, as well as two presidential awards. He was a Fellow of
the AAAS and the APS.

Marcel was born and grew up in Paris. He received a diploma in
comparative western literature from the Sorbonne in 1952, then he moved
to the United States. Remarkably, he began his study of physics only after
enrolling in graduate school at Columbia University.  He received his
PhD under Leon Lederman in 1961. Marcel worked at Columbia for several
years, and was the Deputy Director of the Nevis Laboratory before
coming to the NSF.

Marcel enjoyed telling people, with a twinkle in his eye, that physics
was only his hobby...his real profession was photography. His sensitive
Cibachrome landscapes often explored a deeper vision, revealing a
still, timeless, or mysterious side of reality. His work was widely
exhibited around the world, including two one-man shows at the Corcoran
Gallery of Art in Washington, D.C., as well as exhibits at the Leo
Castelli Gallery in New York and the Troyer, Fitzpatrick, Lassman
Gallery in Washington, D.C.

Marcel's judgment, leadership, wit, and enthusiasm were an inspiration
to all who knew him. He made the world a better place to be in. He will be
missed.

\vfill
\pagebreak

%%%%%%%%%%%%%%%%%%%%%%%%%%%%%%%%%%%%%%%%%%
\section*{\centerline {
Status of the GEO600 project }}
\addtocontents{toc}{\protect\smallskip}
\addtocontents{toc}{\bf Research briefs:}
\addtocontents{toc}{\protect\smallskip}
\addcontentsline{toc}{subsubsection}{\it  
Status of the GEO600 project, by Harald L\"uck}
\begin{center}
    Harald L\"uck, for the GEO team, Universit\"at Hannover, Germany\\
\htmladdnormallink{hal@mpq.mpg.de}
{hal@mpq.mpg.de}
\end{center}
\parindent=0pt
\parskip=5pt

The construction of the GEO600 gravitational-wave detector, a
German-British collaboration of the University of Glasgow, the
Max-Planck-Institut f\"ur Quantenoptik at Garching and Hannover, the
Universit\"at Hannover, the Laser-Zentrum Hannover, the
Albert-Einstein-Institut Potsdam, and the University of Wales College
of Cardiff is well on its way. Short summaries on some aspects of the
current work may give an impression of the status of the project.

\noindent{{\it Civil engineering:}} All the civil engineering work for
the buildings and the clean-rooms was finished about a year ago and
gives a clean room class in the inner sections of the stations of $<
1000$. A clean tent for obtaining class $<$100 clean-room environment
in the vicinity of the tanks is currently being set up.

\noindent{\it Vacuum system:} The vacuum system of 
{{\small GEO}\,600}
approaches
completion. Both 600\,m long tubes are installed and evacuated. One of
them has been baked by passing a DC current of up to 600\,A through
the tube. The tube got an air bake for two days at about
200$^\circ$\,C and a vacuum bake at about 250$^\circ$\,C for a
week. The current pressure in the baked tube is at about
$10^{-8}$\,mbar limited by an air leak which remains to be found. The
second tube where the pressure is still dominated by water vapour is
already thermally insulated and will be baked in a few weeks time.

The vacuum system in the central station can be separated into two
subsystems by a gate valve: 1) The mode-cleaner vacuum system is
installed, evacuated, and baked.  Currently it is disconnected from
the main central cluster and pumped by a separate vacuum pump; 2) The
central cluster, where all the tanks are on site and are currently
being positioned and anchored.

\noindent{\it Mirror suspensions:} The stacks for the mode cleaner
tanks, i.e. three legs per tank of alternating rubber and stainless
steel layers that carry a frame structure from which the mirrors are
suspended, are manufactured and will shortly be installed. The whole
vacuum system of GEO600 is entirely made of metal. To avoid
contamination of the vacuum system by the rubber of these stacks they
are welded into stainless steel bellows that will be pumped
seperately.  The suspensions for the mode cleaner mirrors which will
be hung as double pendulums is halfway through the workshop.

The stacks for those tanks that will contain the main optics will
include an active anti-seismic stage consisting of piezo actuators and
geophones in a digital feedback loop. All the hardware is purchased
and a suppression factor of about 20\,dB has recently been
demonstrated in the sub-Hertz frequency range.

The main optics will be suspended as triple pendulums with a
monolithic fused silica suspension for the lowest stage. Sufficiently
high mechanical Q-factors have been demonstrated by the Glasgow arm of
the GEO600 collaboration. 

\noindent{\it Electronics:} The local control boards are just moving
from the construction into the manufacturing phase. Data communication
via a 6\,Mbit/s radio link to Hannover and from there into the
internet works to our satisfaction. The only difficulty the glass
fibre data transport along the interferometer arms encountered were
mice who in winter time, for the lack of other food, checked whether
glass fibres provide a healthy diet.

\noindent{\it Optics:}  The optical layout of GEO600 has been
changed from external modulation to Schnupp (frontal) modulation hence
avoiding the need of an additional Mach-Zehnder Interferometer at the
Michelson output. In addition the control signal for the differential
Michelson length control and also the control signal for the signal
recycling mirror can be obtained with one modulation frequency by
picking up the signals at different locations in the detector.  This
technique has successfully been demonstrated at the 30\,m prototype in
Garching.

\noindent{\it Laser:}\\ In the time it takes the 10\,W slave laser to
come from a prototype stage to meet the specifications we will use a
stabilised 1\,W Nd-YAG laser, which works well enough to start with.

Additional information about the status and the schedule of 
{{\small GEO}\,600} can be
found in the {{\small GEO}\,600} web pages:
\htmladdnormallink{\protect {\tt http://www.geo600.uni-hannover.de}}
{http://www.geo600.uni-hannover.de}.

\vfill
\pagebreak

%%%%%%%%%%%%%%%%%%%%%%%%%%%%%%%%%%%%%%%%%%
\section*{\centerline {
A Nonperturbative Formulation of String Theory?}}
\addcontentsline{toc}{subsubsection}{\it  
A Nonperturbative Formulation of String Theory?, by Gary Horowitz}
\begin{center}
    Gary Horowitz, UC Santa Barbara\\
\htmladdnormallink{gary@horizon.physics.ucsb.edu}
{gary@horizon.physics.ucsb.edu}
\end{center}
\parindent=0pt
\parskip=5pt

   Last November, a young theorist named Juan Maldacena made a bold
conjecture for a nonperturbative formulation of string theory 
(\htmladdnormallink{hep-th/9711200}{
http://xxx.lanl.gov/abs/hep-th/9711200}).
He considered
spacetimes which asymptotically approach Anti-de Sitter ($AdS$) space
and stated that with this boundary condition, string theory is 
completely equivalent to an ordinary field theory in one less dimension 
which is
conformally invariant i.e. a conformal field theory (CFT). At
first sight, this seems crazy -- but it is not obviously wrong. When the
string theory is weakly coupled and hence well understood,
the field theory is strongly coupled, and
vice versa. It is well known that perturbatively,
string theory has many more degrees of freedom than an ordinary field
theory. However, there have been earlier hints from studies of strings
at high temperature that fundamentally,
string theory should have many fewer degrees of freedom than it appears to.
The conformal boundary  at infinity
of $AdS_n$ is a time-like cylinder  $S^{n-2} \times R$.
For many purposes, one can think of the field theory as living on this
boundary. The most natural field theory to live on this boundary is a 
CFT since the metric is only defined up to conformal transformations.
In a sense, this theory is `holographic' since the fundamental
degrees of freedom live on the boundary, but describe all the physics
taking place inside. 

There are actually
a series of conjectures which differ by the precise asymptotic boundary
conditions one imposes on the spacetime. Perhaps the simplest case
applies to ten dimensional spacetimes which asymptotically approach
$AdS_5\times S^5$. String theory with this boundary condition is conjectured
to be 
 completely equivalent to four dimensional, $U(N)$, ${\cal N}=4$
supersymmetric Yang-Mills theory\footnote{${\cal N}=4$ means that there 
are four
independent supersymmetry transformations. This is the maximum possible
supersymmetry for a gauge theory in four dimensions.}.
The radii of the $S^5$ and
$AdS_5$ are equal (i.e., their scalar curvatures are equal in magnitude,
but of course opposite in sign) and proportional to $N^{1/4}$ in Planck
units.  To describe spacetimes with small curvature asymptotically,
one needs the radii
to be large and hence $N$ to be large.

Maldacena was led to his conjectures by exploring the consequences
of the recent description of quantum states of extreme and near extreme
black holes in string theory. The near horizon geometry of the extreme
Reissner-Nordstr\"om  solution is $AdS_2\times S^2$. In higher dimensions,
the near horizon geometries of  extreme black holes and 
extended black `p-branes'
are also products of $AdS$ and  spheres. It was found a few years ago,
that the quantum states of a near extreme black hole could be described in 
terms of a gauge theory. The gauge theory excitations interacted
with the usual string states which described strings propagating further 
from the
black hole. Maldacena argued that if one takes a certain 
limit which removes the asymptotically flat region around the black hole
and focuses on the near
horizon geometry, the gauge theory completely decouples from the usual
string modes. So it should
describe all the physics of strings propagating near the horizon.

Since we do not have another nonperturbative definition of string theory,
one could simply take the gauge theory as the definition of the 
theory\footnote{The
gauge theory is believed to be well defined non-perturbatively through 
e.g. a lattice regularization.}. However, to 
prove the conjecture, one must show that there is an expansion of the
gauge theory which reproduces the perturbative expansion of string theory
about $AdS_5\times S^5$. More than  twenty years ago, 't Hooft showed that
the $1/N$ expansion of a gauge theory indeed resembles a string theory.
People are now trying to make this correspondence more precise. It is easy
to see that the symmetries agree. The isometry group of
$AdS_5\times S^5$ is $SO(4,2) \times SO(6)$, so perturbative string theory
on this background will have these symmetries. Four dimensional
Yang-Mills theory is invariant under the conformal group $SO(4,2)$. 
The ${\cal N} = 4$ supersymmetry implies that in addition to the gauge
field, there are  four fermions and six scalars, all taking values in the
adjoint of $U(N)$. There is an $SO(6)$ symmetry which rotates the six
scalars, so the bosonic symmetries agree.  It turns out that 
the supersymmetries agree as well. 

The low energy string excitations are
described by a supergravity theory. It has been shown that the energy of
linearized supergravity modes on $AdS_5\times S^5$ agrees precisely with
the energy of states in the gauge theory. This can be verified even though the
gauge theory is strongly coupled, since the supergravity states correspond
to states in the gauge theory which are protected against 
quantum corrections. Some perturbative interactions have also been checked
and shown to agree. The gauge theory is believed to describe {\it all} finite
energy excitations about $AdS_5\times S^5$, including black holes.
It is clear from earlier work  that
the gauge theory has enough states to reproduce the entropy of black holes.
It is a simple exercise to check that a large five dimensional
Schwarzschild-AdS black hole  has $S\propto T^3$ and $M\propto T^4$,
exactly like a four dimensional field theory. 

There is currently a tremendous amount of activity in this area, and the
subject is developing rapidly in many directions. For example,
Witten suggested that one could break supersymmetry, 
and use this conjecture to
study strongly coupled
non-supersymmetric gauge theories 
(\htmladdnormallink{
hep-th/9802150}
{http://xxx.lanl.gov/abs/hep-th/9802150},
\htmladdnormallink{
hep-th/9803131}
{http://xxx.lanl.gov/abs/hep-th/9803131}).
In fact, a simple picture of
confinement is emerging based on the geometry on certain asymptotically
AdS spacetimes. Conversely, efforts are being made to use the
gauge theory to study black hole
evaporation. One immediate consequence seems
to be that the evaporation will be unitary, since the underlying gauge
theory  is unitary and its time is equivalent to the asymptotic
AdS time.
 An application to cosmology has also been suggested 
(G. Horowitz, D. Marolf 
\htmladdnormallink{
hep-th/9805207}
{http://xxx.lanl.gov/abs/hep-th/9805207}).
I have mentioned only a few of the hundreds of papers which have 
appeared. There was also a recent discussion in Physics Today 
(August 1998, p. 20).

\vfill
\pagebreak

%%%%%%%%%%%%%%%%%%%%%%%%%%%%%%%%%%%%%%%%%%
\section*{\centerline {
TAMA project update}}
\addcontentsline{toc}{subsubsection}{\it  
TAMA project update, by Seiji Kawamura}
\begin{center}
Seiji Kawamura, National Astronomical Observatory, Japan \\
\htmladdnormallink{seiji@gravity.mtk.nao.ac.jp}
{seiji@gravity.mtk.nao.ac.jp}
\end{center}
\parindent=0pt
\parskip=5pt

TAMA is the Japanese project of building a gravitational wave
detector. The detector employs a standard power recycled Fabry-Perot
interferometer with an arm length of 300m. It is located on the campus
of National Astronomical Observatory in Tokyo. The project started in
1995 with the collaborating efforts of National Astronomical
Observatory, The University of Tokyo, Institute for Cosmic Ray
Research, Institute for Laser Science, High Energy Accelerator Research
Organization, Yukawa Institute of Theoretical Physics, and Osaka
University. Miyagi University of Education joined the project later.

In 1997 the facility and the vacuum system were completed. The 10W
Nd:YAG laser using the injection locking technique was developed by
Sony Corporation. The ring-cavity mode cleaner was installed and
checked in performance with the 500mW Nd:YAG laser. In 1998 the one arm
cavity consisting of two suspended mirrors was installed and locked
using the 700mW Nd:YAG laser with both mirrors controlled in
orientation by wave front sensing signals (Ward method). We then
installed the other arm cavity, the beam-splitter, and two pick-off
mirrors. We are now trying to lock the Fabry-Perot Michelson
interferometer. In parallel with this effort, we are connecting the 10W
laser to the mode cleaner. In fall 1998 we will combine the 10W laser
and mode cleaner with the Fabry-Perot Michelson interferometer. When we
increase the sensitivity to a reasonable level in spring 1999, we plan
to take data of the detector for one month.  We will then install the
recycling in the interferometer. We expect to have a full sensitivity
of TAMA at the end of 1999.

The mechanism of the excess noise produced in the light transmitted
through the mode cleaner at the RF frequency of the transmitted
sidebands was well understood, and the noise was well suppressed. This
ensures that the excess noise does not preclude the shot noise limited
sensitivity of the interferometer.

The one arm cavity was held locking very stably for ten days with only
several losses of locking. The wave front sensing signals were well
diagonalized for the two mirrors, and used to control the orientation
fluctuation of the mirrors.

The absolute length of the 300m arm cavity was measured with the
accuracy of 1 micrometer using a new technique, which used the phase
modulation sideband transmitted through the arm cavity. The result
showed that the motion of the 300m arm cavity was dominated by peaks of
20 micrometers which occurred regularly everyday. It turned out that
the peaks were caused by pumping the underground water by a hospital
nearby.

We will use a new signal extraction method for the recycled
interferometer. In the conventional way it is difficult to get the
recycling cavity length signal, because the signal is dominated by the
common-mode arm cavity length signal due to its high finesse. It was
found that the signal extraction matrix can be diagonalized by
optimizing the reflectance of the recycling mirror to the sideband
rather than to the carrier. The principle of this method has been
already verified in the 3m prototype.

As for the future project after TAMA, Institute for Cosmic Ray Research
declared that they have adopted the building of a km-class
gravitational wave antenna as the institute's future main project. This
greatly increases the chances for a future km-class antenna in Japan.

More information about TAMA can be found at our web site.
\htmladdnormallink{\protect {\tt http://tamago.mtk.nao.ac.jp/}}
{http://tamago.mtk.nao.ac.jp/}

\vfill
\pagebreak

%%%%%%%%%%%%%%%%%%%%%%%%%%%%%%%%%%%%%%%%%%
\section*{\centerline {
Neohistorical approaches to quantum gravity}}
\addcontentsline{toc}{subsubsection}{\it  
Neohistorical approaches to quantum gravity, by Lee Smolin}
\begin{center}
Lee Smolin, Penn State\\
\htmladdnormallink{smolin@phys.psu.edu}
{smolin@phys.psu.edu}
\end{center}
\parindent=0pt
\parskip=5pt

In the old days of quantum gravity work was split between canonical
and histories approaches.  Since the invention of the Ashtekar
formalism [1] and loop quantum gravity [2,3] more than
ten years ago, much non-string theory work in quantum gravity was
devoted to canonical approaches. The main results of this work have
been the discovery that geometrical quantities such as areas and
volumes [4,5]  are represented by finite operators with
discrete spectra, whose eigenstates give a basis of states which may
be described in terms of spin networks [6,7].  There have
even been rigorous theorems demonstrating that these results must be
true of a large class of quantum theories of
gravity [8]\footnote{Many people have worked in this area.
For the older work I give here only a few representative references
and apologize to all those not cited.}.

In spite of these successes, 
in the last two years there has been a shift back to 
approaches that emphasize spacetime histories and path integrals.
One reason for this has been the realization, following the 
work of Thiemann [9], that while quantum general relativity
may be a finite and well defined theory at the Planck scale, there is
evidence that the theory produced by the method canonical quantization
does not have a continuum limit which reproduces classical
general relativity [10,11].  As a result, the interest
of many people turned to the possibility that the dynamics of the
spin network states might be described in terms of a histories framework
in a way that avoid the difficulties of the canonical approach.

In fact, even before these developments, the first formulation of a path
integral framework to describe the dynamics of spin network states
had been proposed by Reisenberger [12]. The connection
of Reisenberger's proposal to the canonical formalism was then
elucidated in papers with and by Rovelli [13,14].
In their proposal, and
much subsequent work, the spacetime histories in the path integral
are represented as discrete combinatorial structures, as is fitting
as the basis states in the spin network representation are themselves
largely combinatorial.  These combinatorial structures can be
often visualized as four dimensional triangulations, with labels
associated with the discrete geometrical quantities
attached to edges, surfaces and tetrahedra.  However, unlike the
Regge calculus and dynamical triangulation formulations, the triangulations
are not meant as an approximation, but rather as a representation of the
discrete structure of quantum geometry which was revealed by the
results of the canonical formulation.  

One strength of this approach is that it has merged with a set of
developments in mathematics, which were also going on for several
years.  Since the early 1990's Louis Crane and collaborators have
been working on extending topological quantum field theory (TQFT)
from three to four dimensions [15].  The main goal of this work, 
so far 
unrealized, has been to find a combinatorial formulation of the
Donaldson invariant.  In three dimensions, TQFT relies on powerful and
apparently deep connections between topology, combinatorics and
representation theory, which are most succinctly expressed in terms
of category theory.  Crane, Frenkel, Yetter, Baez, Dolan and others have
been exploring the idea that four dimensional TQFT's must involve still
more intricate and subtle relations between the discrete and continuous,
which they refer to as ``extended category theory."  In all of these
theories, a topological invariant is expressed as a sum over labels
assigned to various parts of the triangulation of a manifold.
Topological invariance is proven by showing that the sum is independent
of the choice of triangulation.  Such formulations are called
``state sum models".

The relationship of this work to quantum gravity came about because
several people  had noticed that classically general relativity can be
understood as arising from a TQFT by the imposition of a constraint local
in the fields [16].  Crane thus proposed a program of constructing
quantum gravity as a discrete path integral by imposing an analogous
constraint on a state sum model of a TQFT [17]
This goal has apparently
been realized by a recent proposal of Barrett and Crane [18],
which
has been studied in detail by Baez [19]  and others [20].
Such models of quantum gravity were called by Baez, ``spin foam models"
in homage to the spacetime foam of John Wheeler, and the name seems
to have stuck.  

One thing that is very impressive about these models is that the
amplitudes for the histories do seem to reproduce, history by
history, the Regge action for classical general 
relativity [21],
while being derived by a particularly elegant restriction of an
expression for a topological invariant.  This is good because if such
a theory is to have a continuum limit, it must be special so as to avoid
the problems which prevent generic non-perturbatively-renormalizable
theories from making sense.  The fact that these theories are closely
related to TQFT's, which by definition have continuum limits, suggest that
there is reason to hope that this is the case.

Like the earlier Regge calculus and dynamical triangulation approaches,
these new spin foam models of discrete quantum gravity are, so far,
Euclidean, in that the histories represent discretizations of a
four dimensional manifold of Euclidean signature.  One expects that
if such theories have continuum limits, they are related to second
order equilibrium critical phenomena, of the type searched for in
the older models.

An alternative approach to discrete spacetime histories which is
intrinsically Lorentzian was then proposed by Markopoulou [22]
and versions of it have been studied by several 
people [23,24,25,26].  In these
theories spacetime is a discrete causal set, of the kind studied by
Sorkin, Meyers and collaborators [27], and 
't Hooft [28], with
the additional structure that the causally unrelated sets (discrete
spacelike slices) are closely related to the spin network states.
Thus, the  basic idea of Markopoulou is that spacetime is made of a 
discrete 
set of events which correspond to local changes in the spin network
states.  

The question of the existence of a continuum limit for these kinds of
theories has been studied by Abjorn and Loll in the $1+1$ dimensional
case [24].  General considerations discussed in 
[24]
suggest that
in such theories the continuum limit may be analogous to that found
in certain problems in non-equilibrium critical phenomena such as
directed percolation, where there are analogues of fluctuating causal 
structure.
The relationship of these causal histories to the Euclidean histories
described by spin foam models is also being investigated [25].
The optimistic
expectation here is that there will be a non-perturbative analogue of
Euclidean continuation that will connect the two classes of theories.

Finally, it should be mentioned, that while new, these developments already
show promising links to other approaches to quantum gravity.  It is
easy to extend the algebras that give rise to the labels so as to
incorporate supersymmetry, perhaps giving rise to a non-perturbative
formulation of string theory [26,29].  Perturbations 
of these
discrete histories indeed look something like perturbative 
strings [30].
There are also possibilities that the special forms of the theory
may allow the holographic principle to be formulated intrinsically,
exploiting the fact that the TQFT's already naturally give finite
dimensional state spaces associated with boundaries of 
spacetime [17]
Finally, the categorical framework of the spin foam models may allow
connections to be made [31]  to the Gell-Mann Hartle, kind of 
histories
formulations [32] as these have also been formulated categorically
by Isham [33].

In general the most important feature of these theories may be that they 
depend
on the deep connections between the continuous and discrete, and between
the topological and algebraic, expressed in categorical terms in
TQFT.  These give a possibility of joining relativity and quantum theory
at their roots, in a way that addresses both the technical and
conceptual questions that have so far stymied all attempts to make a
complete background independent formulation of quantum gravity.

References:
\parskip=2pt	
[1] A. Ashtekar, Phys. Rev. Lett. 57 (1986) 2244;
Phys. Rev. D36 (1987) 1587.

[2] T. Jacobson and L. Smolin,  Nucl. Phys.
B (1988)

[3] C. Rovelli and L. Smolin, Phys Rev Lett 61 (1988) 1155; Nucl 
Phys B133, 80 (1990)

[4] L  Smolin: in {\it Quantum Gravity and 
Cosmology}, eds  J  P\'erez-Mercader {\it et al}, World Scientific, 
Singapore 1992. 

[5] C. Rovelli and L. Smolin
 Nuclear Physics B 442 (1995) 593.  Erratum: Nucl. Phys.
B 456 (1995) 734.

[6] R. Penrose,  
{\it Theory of quantized directions} unpublished manuscript.; 
in {\it Quantum theory and 
beyond}  ed T Bastin, Cambridge U Press 1971;
in {\it Advances in Twistor Theory}, ed. L. P. Hughston and R. S. 
Ward,
(Pitman,1979) p. 301; in {\it Combinatorial Mathematics and
its Application} (ed. D. J. A. Welsh) (Academic Press,1971).

[7] C. Rovelli and L. Smolin,  
\htmladdnormallink{gr-qc/9505006}
{http://xxx.lanl.gov/abs/gr-qc/9505006}, Physical Review D 52 (1995) 5743-5759.

[8] A Ashtekar J Lewandowski D Marolf J 
Mour\~{a}o T Thiemann, \htmladdnormallink{gr-qc/9504018}
{http://xxx.lanl.gov/abs/gr-qc/9504018}, JMP 36 (1995) 519;
A. Ashtekar and J. Lewandowski, \htmladdnormallink{gr-qc/9602046}
{http://xxx.lanl.gov/abs/gr-qc/9602046}; 
J. Lewandowski, \htmladdnormallink{gr-qc/9602035}
{http://xxx.lanl.gov/abs/gr-qc/9602035}.

[9] T. Thiemann,  \htmladdnormallink{gr-qc/9606089}
{http://xxx.lanl.gov/abs/gr-qc/9606089}; 
\htmladdnormallink{gr-qc/9606090}
{http://xxx.lanl.gov/abs/gr-qc/9606090}.

[10] L. Smolin, 
\htmladdnormallink{gr-qc/9609034}
{http://xxx.lanl.gov/abs/gr-qc/9609034}.

[11] R. Gambini, J. Lewandowski, D. Marolf, J. Pullin,
\htmladdnormallink{gr-qc/9710018}
{http://xxx.lanl.gov/abs/gr-qc/9710018}.

[12] M. Reisenberger, 
\htmladdnormallink{gr-qc/9711052}
{http://xxx.lanl.gov/abs/gr-qc/9711052}.

[13] M. Reisenberger and C. Rovelli,
\htmladdnormallink{gr-qc/9612035}
{http://xxx.lanl.gov/abs/gr-qc/9612035}, 
Phys.Rev. D56 (1997) 3490-3508.

[14] Carlo Rovelli, 
\htmladdnormallink{gr-qc/9806121}
{http://xxx.lanl.gov/abs/gr-qc/9806121}

[15] L. Crane and D. Yetter, {\it On algebraic
structures implicit in topological quantum field theories},
Kansas preprint, (1994); in {\it Quantum Topology}
(World Scientific, 1993) p. 120; 
L. Crane and I. B. Frenkel, J. Math. Phys.
35 (1994) 5136-54; J. Baez,   
\htmladdnormallink{q-alg/9705009}
{http://xxx.lanl.gov/abs/q-alg/9705009}.

[16] R. Capovilla, J. Dell and T. Jacobson,
Phys. Rev. Lett. 21 (1989) 2325; Class. Quant. Grav. 8 (1991) 59; 
R. Capovilla, J. Dell, T. Jacobson and L. Mason, Class
and Quant. Grav. 8 (1991) 41.

[17] L. Crane, {\it Clocks and Categories, is
quantum gravity algebraic?} to appear in the special issue of
J. Math. Phys. on quantum geometry,  {\it Categorical Physics},
in {\it Knot theory and
quantum gravity} ed. J. Baez, (Oxford University Press).

[18] J. Barrett and L. Crane, 
\htmladdnormallink{gr-qc/9709028}
{http://xxx.lanl.gov/abs/gr-qc/9709028} 
 
[19] J. Baez, 
\htmladdnormallink{gr-qc/9709052}
{http://xxx.lanl.gov/abs/gr-qc/9709052}.
 
[20]  L. Freidel, K. Krasnov, hep-th/9804185, hep-th/9807092.

[21] L. Crane, D.N. Yetter, 
\htmladdnormallink{gr-qc/9712087}
{http://xxx.lanl.gov/abs/gr-qc/9712087};
L. Crane,  gr-qc/9710108; J. Barrett and R. Williams,
in preparation.

[22] F. Markopoulou,   
\htmladdnormallink{gr-qc/9704013}
{http://xxx.lanl.gov/abs/gr-qc/9704013}.

[23] F. Markopoulou and L. Smolin 
\htmladdnormallink{gr-qc/9702025}
{http://xxx.lanl.gov/abs/gr-qc/9702025}, 
Nucl.Phys. B508, 409-430 (1997). 

[24] J. Ambjorn, R. Loll,hep-th/9805108;
 J. Ambjorn, J.L. Nielsen, J. Rolf, R. Loll,
 hep-th/9806241.

[25] S. Gupta; R. Borissov and S. Gupta, in preparation.

[26] F. Markopoulou and L. Smolin 
\htmladdnormallink{gr-qc/9712067}
{http://xxx.lanl.gov/abs/gr-qc/9712067}; 
hep-th/9712148. 

[27] L. Bombelli, J. Lee, D. Meyer and
R. D. Sorkin,   Phys. Rev. Lett. 59 (1987) 521.

[28] G. 't Hooft,  {\it Quantum gravity: a fundamental 
problem and some
radical ideas.}  Carg\`{e}se Summer School Lectures 1978. 
Publ. ``Recent
Developments in Gravitation''.  Carg\`{e}se 1978. 
Ed. by M. L\'{e}vy
and S. Deser.  Plenum, New York/London, 323;
NATO lectures 
\htmladdnormallink{gr-qc/9608037}
{http://xxx.lanl.gov/abs/gr-qc/9608037}; 
J. Mod. Phys. A11 (1996) 
4623-4688 
\htmladdnormallink{gr-qc/9607022}
{http://xxx.lanl.gov/abs/gr-qc/9607022}.

[29] L. Crane, 
\htmladdnormallink{gr-qc/9806060}
{http://xxx.lanl.gov/abs/gr-qc/9806060}.

[30] L. Smolin, {\it Strings from perturbations
of causally evolving spin networks} preprint, Dec. 1997.

[31] F. Markopoulou, {\it A universe of
partial observers}, (in preparation).

[32] Murray Gell-Mann, James B. Hartle, 
\htmladdnormallink{gr-qc/9404013}
{http://xxx.lanl.gov/abs/gr-qc/9404013}; 
J. B. Hartle, 
\htmladdnormallink{gr-qc/9808070}
{http://xxx.lanl.gov/abs/gr-qc/9808070}.

[33] C. J. Isham, Int.J.Theor.Phys. 36 (1997) 785-814,
\htmladdnormallink{gr-qc/9607069}
{http://xxx.lanl.gov/abs/gr-qc/9607069}.

\parskip=5pt
\vfill
\pagebreak

%%%%%%%%%%%%%%%%%%%%%%%%%%%%%%%%%%%%%%%%%%
\section*{\centerline {
LIGO Project update}}
\addcontentsline{toc}{subsubsection}{\it  
LIGO Project update, by David Shoemaker}
\begin{center}
David Shoemaker, LIGO-MIT\\
\htmladdnormallink{dhs@ligo.mit.edu}
{dhs@ligo.mit.edu}
\end{center}
\parindent=0pt
\parskip=5pt

The Hanford (Washington) and Livingston (Louisiana) LIGO sites are now
really the centers of activity for the LIGO Laboratory. The buildings and
infrastructure are complete, and both permanent and visiting staff are
rapidly on the rise. 

The physical installation of the beam tubes is now complete, with all 8km
accepted at Hanford, 4km accepted at Livingston, and only 4km to go. The
LIGO Lab has started the Hanford ``bakeout'' in which lengths of the over 1m
stainless steel diameter tube is heated by passing current through it 
($I^2R$
works!) to drive out residual gas; this is a huge undertaking, with
monstrous power supplies, cabling, and power sub-stations which travel from
arm to arm and then to Livingston as we bake out the beam tube by 2km
sections. 

The vacuum equipment which houses the detector itself is now all on-site
and in various stages of installation and testing. There are some
complications with the large gate valves which segment and separate the
beam tube from the vacuum equipment, with research into rubber and bellows
and the usual details that make up the whole. This may delay the
availability of the entire vacuum system, but there is lots to do that is
independent and no significant impact on LIGO's turn-on is anticipated. 

Several exciting milestones for the detector have taken place over the
last half-year. The seismic isolation stack first article was
installed and tested, allowing checks of the fit of the components and
a verification of the filtering of seismic noise. Some lessons were
learned both in the process of the installation and some manufacturing
details which can now be applied to the whole series of isolation
systems to follow. The manufacture of the series isolators, and the
installation of the structures, is now underway at Hanford.

Another exciting step is the appearance of the Pre-Stabilised Laser
equipment (and team!) at the Hanford site. The laser, which has already
been built up and tested at Caltech, will be resurrected after shipping
over the next few months, and will be turned on this fall. 

Many other parts of the detector are turning from dreams into hardware,
with optics, suspensions, coupling telescopes, mode-cleaning cavities, and
servo systems all deep into fabrication. High-speed computer backbones are
running at Hanford. The first integration of multiple subsystems will take
place in Fall 98, with light coupled from the Laser into the Input Optics
system which the University of Florida is delivering and installing. 

The MIT LIGO Lab has moved, in July, from its familiar and favorite
quarters in ramshackle Building 20 to a sparkling new space in NW17;
it is equipped with a beautiful high-bay space (for a to-be-installed
full-scale test interferometer), clean-room air everywhere, and (can
you imagine it) carpeting on the office floors. Not so many mice, or
much ``charm'', yet, but a very nice space. The Systems Integration
effort including the data analysis group at Caltech has moved also
into new quarters in the thin air at the top of the Millikan Library.

Research and Development for the initial detector is finishing up. The
work on fringe-splitting at MIT concluded with a test of a digital
servo-loop, confirming our ability to control and perform diagnostics
with the required dynamic range. At Caltech, the 40m interferometer is
allowing tests of alignment and acquisition systems and models.

Research for the next phase of LIGO is heating up. The LIGO Science
Collaboration (LSC) now has significant momentum, and planning for data
analysis of the initial LIGO run and for the hardware improvements slated
for some five years from now are well underway. The last meeting of the
LSC, in August, was held at JILA in Boulder, Colorado; the next meeting is
to be held at the University of Florida in March '99. 

Our schedule calls for shakedown of the interferometers starting in
mid-'99, and operation in 2001. Additional information about LIGO,
including our newsletter and information about the LSC, can be
accessed through our WWW home page at 
\htmladdnormallink{\protect {\tt http://www.ligo.caltech.edu}}
{http://www.ligo.caltech.edu}.

\vfill
\pagebreak
\section*{\centerline {
Gravitational waves from neutron stars:} \\
\centerline{recent developments}}
\addcontentsline{toc}{subsubsection}{\it 
Gravitational waves from neutron stars: recent developments,
by \'Eanna Flanagan}
\begin{center}
    \'Eanna Flanagan, Cornell \\
\htmladdnormallink{eef3@cornell.edu}
{mailto:eef3@cornell.edu}
\end{center}
\parindent=0pt
\parskip=5pt

\def\agt{
\mathrel{\raise.3ex\hbox{$>$}\mkern-14mu\lower0.6ex\hbox{$\sim$}}
}
\def\alt{
\mathrel{\raise.3ex\hbox{$<$}\mkern-14mu\lower0.6ex\hbox{$\sim$}}
}

The gravitational waves that bathe the Earth presumably do not vary
wildly in strength from year to year.  However, our very imperfect
understanding of their strengths occasionally does change abruptly,
prompted either by new astrophysical observations or by new theoretical
predictions and discoveries.  Such has been the case in the last year
for our expectations for periodic gravitational waves from
neutron stars, in two different scenarios: (i) accreting neutron stars in
low mass X ray binaries (LMXBs), and (ii) hot, young neutron stars in
the first year or so after their formation.

\medskip
\noindent
{\it Accreting neutron stars in LMXBs:}

As is well known, rotating neutron stars can radiate via three
mechanisms: (i) non-axisymmetry of the star, (ii) non-alignment
between the axis of rotation and a principle axis of the moment of
inertia tensor, and (iii) excitation of the stars' normal modes.  For
mechanism (i), the amount of non-axisymmetry can be parameterized by
the equatorial eccentricity $\varepsilon_e = (I_{xx}-I_{yy})/I_{zz}$,
where $I_{ij}$ is the moment of inertia tensor and the rotation axis
is the $z$ axis.  Advanced LIGO interferometers can see
non-axisymmetric neutron stars out to $\sim 1$ kpc for $\varepsilon_e
\agt 10^{-7} / f_{500}$ with 1/3 year integration time, where the
rotation frequency is $500 f_{500}$ 
Hz (Thorne 1998).  The likely values of $\varepsilon_e$ for various
neutron star populations are highly uncertain.  Aside from the
millisecond pulsar population which is highly constrained, we know only that
$\varepsilon_e \alt 10^{-5}$ (Thorne 1998).

Lars Bildsten has recently given fairly convincing theoretical and
observational arguments that many LMXBs like Scorpius X-1 should have
values of $\varepsilon_e$ of order $10^{-7}$ or larger and should thus
be fairly strong sources (Bildsten 1998).
First, recent observations by the Rossi X-Ray Timing Explorer
satellite indicate that many of the rapidly accreting stars have spin
frequencies clustered near $300$ Hz.  This is somewhat of a puzzle
since the accretion would be expected to spin up the stars to much
higher frequencies.  Bildsten suggests an explanation for this puzzle:
that gravitational wave emission is preventing these sources from
being spun up any further, i.e., that all the angular momentum being
accreted is being radiated into gravitational waves.  The limiting
angular velocity then scales as the $1/5$th power of $\varepsilon_e$
and is thus fairly insensitive to the amount of non-axisymmetry.
Second, Bildsten suggests a specific mechanism for generating the required
non-axisymmetry: that lateral temperature
gradients due to non-uniform accretion over the surface of the star lead (via
temperature-dependent electron capture reactions) to lateral density
variations in the crust.  The resulting estimated values of
$\varepsilon_e$ are of the order $10^{-7}$, consistent with what is
required.  

The wave strengths for these sources can be predicted directly from
the the observed X-ray flux and the inferred accretion rate; the
amount of non-axisymmetry (quadrupole) is determined by demanding
equality of the spin-up and spin-down torques.  The strongest source,
Sco X-1, is predicted to be detectable with $\sim 3$ years integration
with initial LIGO interferometers (Bildsten 1998).  Thus, a priority
for the early data runs for LIGO and also VIRGO and GEO will be
directed searches for periodic signals from known accreting neutron
stars.

\medskip
\noindent
{\it Hot young neutron stars -- the \mbox{{\rm r}}-mode instability:}

Just over a year ago, Andersson discovered that the
$r$-modes of rotating neutron stars are unstable in the absence of
viscosity, for all values of the star's angular velocity
(Andersson 1998, see also Friedman and Morsink 1998).
The instability is driven by gravitational radiation via the
Chandrasekhar-Friedman-Schutz (CFS) mechanism (Chandrasekhar 1970, Friedman
and Schutz 1978), and was reviewed by Sharon Morsink in 
Issue 10 of Matters of Gravity (Morsink 1997).
Over the past year a flurry of papers have explored
the dramatic astrophysical consequences of the $r$-mode instability.  In
this review I'll describe these predicted consequences, 
and summarize the uncertainties and implications.  [For a detailed
review of instabilities in rotating stars see Stergioulas 1998].

The picture that is emerging is the following.  When a neutron star is
first formed it is likely spinning at a substantial fraction of its
maximum angular 
velocity.  While the star cools from $\sim 10^{11} K$ to $\sim 10^{9}
K$ via neutrino emission over the first few years of its life, the
stars' $r$-modes are excited and radiate copious amounts of
gravitational radiation, carrying away as much as $0.01 M_\odot c^2$ of
energy and most of the initial angular momentum of the star
(Lindblom et al. 1998, Andersson et al. 1998).  When the
transition to a superfluid state occurs at $T \sim 10^9 K$, the star
is left with angular velocity of $\Omega = (0.05 - 0.10) \Omega_{\rm 
max}$, the exact value being somewhat uncertain.  Here $\Omega_{\rm
max}$ is the maximum allowed angular velocity.  The predicted wave
strengths are such that these sources could be seen out to the VIRGO
cluster ($r \sim 20 \ {\rm Mpc}$) with enhanced LIGO interferometers
(Owen et al. 1998).  This is quite an exciting prospect since the event
rate could be many per year.

This scenario is consistent with the inferred spin after formation of
the Crab pulsar of about $0.05 \Omega_{\rm max}$, and also (within the
uncertainties of the predictions) of the initial spin period of $\sim
7 \, {\rm ms}$ of the recently discovered young pulsar PSR J0537-6910
(Owen et al. 1998).  It also resolves the observational puzzle that
neutron stars seem to be formed with rather small spins despite one's
expectation of near maximal initial spins due to conservation of angular
momentum during stellar core collapse.  It rules out
accretion-induced-collapse of white dwarfs as a mechanism for forming
millisecond pulsars; millisecond pulsars must form instead via accretion in which
the temperature never gets hot enough to trigger the $r$-mode
instability.  Finally,  since it now seems more likely than before that
typical stellar core collapses involve rapid rotation rates, it
improves the prospects of our detecting supernovae.

Turn now to the assumptions and calculations that underlie these
predictions.  There 
are two conditions for a mode in a realistic neutron star 
to be CFS-unstable:  (i) The mode must be retrograde with  
respect to the star but prograde with respect to distant inertial
observers, the classical CFS condition.  Not all $r$-modes will
satisfy this condition (Lindblom and Ipser 1998), but 
the dominant $l=m=2$ $r$-mode will do so.  
A crucial point is that this condition is satisfied for
all values of $\Omega$ for unstable $r$-modes, whereas it is only
satisfied at large $\Omega$ for the previously considered $f$-modes.
(ii) When one measures the mode's energy in the rotating frame, the
amount of energy 
lost to viscous dissipation per cycle must be less than the amount of
energy per cycle that gravitational radiation reaction adds to the
mode.  In other words, the viscous dissipation timescale must be longer
than the instability growth timescale.  For the original CFS
instability, calculations in Newtonian and
post-Newtonian gravity (Lindblom 1995) had shown that these
conditions are satisfied for the $l=m=2$ $f$-mode only in a certain region
in the $T - \Omega$ plane (where $T$ is the stellar temperature) with
$\Omega \agt 0.9 
\Omega_{\rm max}$ and $10^9 K \alt T \alt 10^{10} K$,
where $\Omega_{\rm max}$ is the maximum angular velocity.  The
dependence on temperature arises due to the strong dependence of the
coefficients of bulk and shear viscosity on temperature.   
Since neutron stars are at temperatures $\agt 10^9 K$ only for the
first few years after their formation, and since it was not clear that
the initial value of $\Omega$ would be $\agt 0.9 \Omega_{\rm max}$, 
the conventional view was that the CFS instability would probably not
be important in practice (see, eg, Thorne 1998).
This picture changed when the $r$-mode instability was discovered.  
Two independent calculations in Newtonian gravity using a slow
rotation approximation have indicated that 
the instability region in the $T - \Omega$
plane is much larger for $r$-modes, extending down to $\Omega \agt
0.05 \Omega_{\rm  max}$ (Lindblom et al.\ 1998, Andersson et al.\
1998).  

For a newly born neutron star, the evolution of the
stars angular velocity and of the $r$-mode amplitude was solved for by
making the following assumptions (Owen et al. 1998): Assume that only the
dominant, $l=m=2$, $r$-mode is relevant.  Assume that the mode
amplitude grows due to the instability until it saturates at a
value of order unity due to nonlinear effects.  [The predictions are
not very sensitive to the assumed saturated value of the mode amplitude].
Then, use conservation of angular momentum to solve for the spin
down of the star.  While these assumptions seem reasonable, it will
be important to verify the qualitative predictions by numerical
calculations that allow for nonlinear mode-mode couplings, perhaps
using post-Newtonian hydrodynamic codes.  There are also uncertainties
related to the values of the viscosity coefficients and the
temperature of the transition to superfluidity; investigation into
these issues is continuing.  However, the overall picture of rapid
spindown seems very robust.  

To conclude, it is not often that elegant but somewhat arcane 
aspects of general relativity (like radiation reaction due
to current multipoles) have such dramatic astrophysical and
observational consequences.  Let us hope for many more such
discoveries.

\bigskip

{\bf References:}

Andersson, N., 1998, Ap.\ J.\ {\bf 502}, 708A (also 
\htmladdnormallink{gr-qc/9706075}
{http://xxx.lanl.gov/abs/gr-qc/9706075}).\hfil\break
% A new class of unstable modes of rotating relativistic stars
%
Andersson, N., Kokkotas, K., and Schutz, B.F., 1998, 
%{\it Gravitational radiation limit on the spin of young neutron stars},
\htmladdnormallink{astro-ph/9805225}
{http://xxx.lanl.gov/abs/astro-ph/9805012}.\hfil\break    
Bildsten, L., 1998, Ap.\ J.\ {\bf 501}, L89 (also 
\htmladdnormallink{astro-ph/9804325}
{http://xxx.lanl.gov/abs/astro-ph/9804325}).\hfil\break    
% GRAVITATIONAL RADIATION AND ROTATION OF ACCRETING NEUTRON STARS
%
Chandrasekhar, S., 1970,  Phys. Rev. Lett., {\bf 24}, 611.\hfil\break
%
%Cutler, C. and Lindblom, L., 1987, Ap.\ J., {\bf 314} 234. \hfill\break
%
Friedman, J.L., 1978, Commun. Math. Phys., {\bf 62} 247.\hfil\break
Friedman, J.L. and Morsink, S.M., 1998, Ap. J. {\bf 502}, 714 (also 
\htmladdnormallink{gr-qc/9706073}
{http://xxx.lanl.gov/abs/gr-qc/9706073}).\hfil\break  
% AXIAL INSTABILITY OF ROTATING RELATIVISTIC STARS
%
Friedman, J.L. and Schutz, B.F., 1978, Ap.\ J, {\bf 222}, 281.\hfil\break
%
%Hiscock, W.A., 1998, 
%{\it Gravitational waves from rapidly rotating white dwarfs}, 
%\htmladdnormallink{gr-qc/9807036}
%{http://xxx.lanl.gov/abs/gr-qc/9807036}.\hfil\break    
%
%Kojima, Y., 1998, M.N.R.A.S. {\bf 239}, 49 (also 
%\htmladdnormallink{gr-qc/9709003}
%{http://xxx.lanl.gov/abs/gr-qc/9709003}).\hfil\break  
% Quasi-toroidal oscillations of rotating relativistic stars
%
%Kokkotas, K.D. and Stergioulas, N., 1998, 
%{\it Analytic description of the r-mode instability in uniform density stars},
%\htmladdnormallink{astro-ph/9805297}
%{http://xxx.lanl.gov/abs/astro-ph/9805297}.\hfil\break   
%
Lindblom, L., 1995, Ap.\ J.\ {\bf 438}, 265.\hfil\break   
Lindblom, L. and Ipser, J.R., 1998, 
%{\it The r-modes of the Maclaurin Spheroids}, 
\htmladdnormallink{gr-qc/9807049}
{http://xxx.lanl.gov/abs/gr-qc/9807049}.\hfil\break    
Lindblom, L., Owen, B.J., and Morsink, S.M., 1998, Phys.\ Rev.\ Lett.\
{\bf 80}, 4843 (also \htmladdnormallink{gr-qc/9803053}
{http://xxx.lanl.gov/abs/gr-qc/9803053}).\hfil\break  
%
%Madsen, J., 1998, 
%{\it How to identify a Strange Star},
%\htmladdnormallink{astro-ph/9806032}
%{http://xxx.lanl.gov/abs/astro-ph/9806032}.\hfil\break    
%
Morsink, S.M., 1997, {\it Instability of rotating stars to axial
perturbations}, MOG No.\ 10, (also 
\htmladdnormallink{gr-qc/9709023}
{http://xxx.lanl.gov/abs/gr-qc/9709023}.\hfil\break    
%
%Morsink, S.M., Stergioulas, N., and Blattnig, S.R., 1998, 
%{\it Quasi-normal modes of rotating relativistic stars --- neutral modes
%for realistic equations of state}, 
%\htmladdnormallink{gr-qc/9806008}
%{http://xxx.lanl.gov/abs/gr-qc/9806008}.\hfil\break    
%
Owen, B.J., Lindblom, L., Cutler, C., Schutz, B.F., Vecchio, A., and 
Andersson, N., 1998, 
%{\it Gravitational waves from hot young rapidly rotating neutron
%stars},
\htmladdnormallink{gr-qc/9804044}
{http://xxx.lanl.gov/abs/gr-qc/9804044}.\hfil\break  
%
%Papaloizou J. and Pringle, J.E., 1978, MNRAS, {\bf 182} 423.
%\hfil\break
%
Stergioulas, N., 1998, 
%{\it Rotating Stars in Relativity}, 
Living Reviews in Relativity, Vol 1 (also \htmladdnormallink{gr-qc/9805012}
{http://xxx.lanl.gov/abs/gr-qc/9805012}).\hfil\break  
Thorne, K.S., 1998, In R.~M. Wald, editor, {\it Black Holes and Relativistic
Stars}, pages 41-77, University of Chicago Press (also 
\htmladdnormallink{gr-qc/9706079}
{http://xxx.lanl.gov/abs/gr-qc/9706079}).\hfil\break  
%
%Thorne, K.S. and Campolattaro, A., 1967, Ap.\ J, {\bf 149} 591.\hfil\break 
%

\vfill
\pagebreak

\vfill

\pagebreak

\section*{\centerline {Perugia Meeting}}
\addtocontents{toc}{\protect\smallskip}
\addtocontents{toc}{\bf Conference Reports:}
\addtocontents{toc}{\protect\smallskip}
\addcontentsline{toc}{subsubsection}{\it  Perugia Meeting, by Joe Kovalik}
\begin{center}
    Joe Kovalik, Perugia Group, VIRGO Project\\
\htmladdnormallink{Joe.Kovalik@pg.infn.it}
{Joe.Kovalik@pg.infn.it}
\end{center}
\parindent=0pt
\parskip=5pt

The Meeting on Thermal Noise and Low Frequency Noise Sources in
Gravitational Wave Detectors took place in Perugia, Italy, on June
4-6.

There has recently been a significant growth of interest in low frequency
noise sources in gravitational wave detectors and in particular, thermal
noise.  This workshop was organized to bring together various international
efforts in this field and to discuss current results openly in a workshop
format.

All the major international groups were present including:  VIRGO, GEO600,
LIGO, LIGO science community affiliated groups, TAMA, ACIGA, AURIGA, and
Nautilus.  The meeting started with a series of review talks on thermal noise.

The rest of the meeting was organized into three main sessions: modelling of
thermal noise and low frequency noise sources, materials and geometries for
low noise detectors and finally advanced designs.

An informal proceedings will be put together at the end of August and
should be available in the fall.  More information is available at
\htmladdnormallink{\protect {\tt http://www.pg.infn.it/virgo/PerugiaWorkshop}}
{http://www.pg.infn.it/virgo/PerugiaWorkshop}.  The main conclusions
of the talks and discussions will be given here.

It is quite clear that most of the groups agree on models, measurements and
predictions.  Moreover, it is also evident that traditional designs based on
fused silica test masses with metallic suspension wires are reaching their
theoretical limits in experimental tests.  Unfortunately, these numbers do not
give excellent results.  The best Q's for fused silica test masses are no more
than 2e7 while typically they can be 5e5.  Sapphire shows some promise of
higher Q's with preliminary measurements in Perth giving Q's of a few 1e7, but
hopefully going up to 1e8. This coupled with the higher frequency internal
resonances of sapphire make it a better material from a thermal noise
performance viewpoint.  There still remains, however, the problems of the
optical properties of sapphire.

For the pendulum mode, it seems that traditional designs cannot give Q's
better than about 1e6.  Fused silica suspension wires do give better results
with some experiments showing Q's better than 1e7.  Unfortunately, this still
is not good enough.  Fused silica does give an immediate design improvement
without going to drastic design changes.

Advanced ideas do offer some hope, but with a few precautions.  There are
proposals for electrostatic and magnetic suspensions.  Many of the bar groups,
however, have shown that there are diverse electrical and magnetic loss
mechanisms and noise sources that must be considered beforehand.  They could
significantly affect the performance of these alternative designs. Active
suspensions are also a solution for reducing seismic noise, but they require
more effort in order to produce fully functional prototypes. 

The final solution is to cool the suspension system.  While it is not
obviously clear that the test mass where laser power is dissipated can be
cooled, the suspensions could be cooled to a few kelvin.  This offers two
advantages.  The thermal noise is reduced by a factor of the square root of
temperature.  Also, many materials (but not fused silica) have better Q's at
low temperature.  There is only some preliminary ideas of what research should
be performed.  This could be a good meeting point for the interferometer
community to learn from the bar community with its many years experience in
cryogenic, low noise experiments.

\vfill
\pagebreak
%%%%%%%%%%%%%%%%%%%%%%%%%%%%%%%%%%%%%%%%%%
\section*{\centerline {
Nickel and Dime gravity meeting}}
\addcontentsline{toc}{subsubsection}{\it  
Nickel and Dime gravity meeting, by Eric Poisson}
\begin{center}
   Eric Poisson, University of Guelph, Canada\\
\htmladdnormallink{poisson@physics.uoguelph.ca}
{poisson@physics.uoguelph.ca}
\end{center}
\parindent=0pt
\parskip=5pt

The second Eastern Gravity Meeting (dubbed Nickel and Dime by the 
organizers, and the sequel to the 1996 New Voices in Relativity 
meeting [1]) was held at the University of Syracuse on March 
28 and 29, 1998. The meeting was hosted by the Syracuse relativity 
group (Josh Goldberg, Don Marolf, Peter Saulson, and Rafael Sorkin) 
and most of the organizing was done by post-doctoral fellow Steve 
Penn. The format of the meeting was identical to the Midwest and 
Pacific Coast meetings, with 15 minutes given to every speaker. 

The meeting brought together approximately 50 participants, of which
approximately half were graduate students. The topics covered ranged
widely, including numerical relativity (Pablo Laguna, Roberto Gomez, 
Pedro Marronetti, Grant Mathews), quantum field theory in curved spacetime 
(Eanna Flanagan, Larry Ford, Ted Jacobson, Wolfgang Tichy), experimental 
LIGO physics (Peter Csatorday, Gabriela Gonzalez, Andri Gretarsson, Ryan 
Lawrence, Steve Penn, Peter Saulson, Bill Startin), gravitational waves 
(Nils Andersson, Serge Droz, Kostas Kokkotas, Eric Poisson), classical 
general relativity (Arley Anderson, Simonetta Frittelli, Thomas Kling, 
Bill Laarakkers), and quantum gravity (Chris Beetle, Ivan Booth, Roumen 
Borissov, Richard Epp, Sameer Gupta, Eli Hawkins, Jim Javor, Kirill 
Krasnov, Jorge Pullin, David Rideout, Rob Salgado, Wendy Smith,
Rafael Sorkin, Sachin Vaidya, Chun-hsien Wu). 

Here are some of the conference's highlights. The following discussion
is necessarily biased toward those topics I am most familiar with, and 
I'm afraid it will not do justice to the many good talks on quantum 
gravity.

Grant Mathews and Pedro Marronetti (Notre Dame) reported on the current
status of their numerical work (carried out with Jim Wilson) on close 
neutron-star binaries [2]. This work has generated quite a bit of 
controversy over the last couple of years. (See Ref. [3] for many 
negative papers, and Ref. [4] for positive contributions.) Mathews, 
Marronetti, and Wilson predict that the central density of the neutrons 
stars increases as the stars approach each other, sufficiently so that the 
stars will undergo gravitational collapse when they are still widely separated. 
This conclusion goes against physical intuition (and the many rebuttals listed
in Ref. [3]) which suggests that the stars' mutual tidal interaction
should make them {\it more stable} against gravitational collapse. In the 
last several months, these authors have been testing their code for possible
errors and inconsistencies, and have found none. It will be interesting to
see how this all gets resolved in the future.

Nils Andersson (T\"ubingen) reported on his recent discovery of the
$r$-mode instability of rotating neutron stars [5]. This instability 
results in the rapid spin-down of a young neutron star, and produces a 
large amount of gravitational waves. Needless to say, this is very 
exciting, because of the possibility that these waves could be 
eventually detected by LIGO/VIRGO/GEO/TAMA. 

Another interesting presentation was from Ted Jacobson (Maryland), who 
discussed the relevance of super-Planckian frequencies in the usual 
derivation of the Hawking effect. He presented recent work carried 
out with Steven Corley [6], in which a quantum scalar field is 
put on a one-dimensional lattice in a black-hole spacetime. The lattice
provides a natural cutoff in the field's dispersion relation, so that
super-Planckian frequencies do not occur. For wavelengths which are long 
compared with the cutoff, the usual Hawking spectrum is recovered.

Finally, Larry Ford (Tufts) reported on an interesting possibility
of amplifying the usual Casimir effect by changing the frequency
spectrum of the contributing field modes, which can be done by 
introducing a dielectric into the problem. As a result of his
analysis, Ford finds that a small dielectric sphere in the vicinity
of a conducting boundary would undergo a potentially measurable force 
which oscillates (between attraction and repulsion) as a function of 
the distance from the boundary.  

Overall this was a marvelous meeting, during which students and postdocs 
met with their peers and talked shop, while their advisors exchanged gossip. 
Like the Pacific Coast and Midwest meetings, the Eastern Gravity Meeting 
provides a friendly and informal forum for all workers in gravitational 
physics. Most importantly, it provides a unique opportunity for graduate
students and postdocs to gain experience at giving talks. It serves a very 
useful purpose, and I wish it a very long life!

References:

[1] L. Smolin, {\it Matters of Gravity} {\bf 9} 
\htmladdnormallink
{(Spring 1997)}
{http://vishnu.nirvana.phys.psu.edu/mog9/mog9.html}.

[2] P. Marronetti, G.J. Mathews, and J.R. Wilson 
\htmladdnormallink
{gr-qc/9803093}
{http://xxx.lanl.gov/abs/gr-qc/9803093}; 
\htmladdnormallink
{gr-qc/9710140}
{http://xxx.lanl.gov/abs/gr-qc/9710140}; 
\htmladdnormallink
{gr-qc/9601017}
{http://xxx.lanl.gov/abs/gr-qc/9601017}.

[3] T.W. Baumgarte, G.B. Cook, M.A. Scheel, S.L. Shapiro, and 
S.A. Teukolsky 
\htmladdnormallink{gr-qc/9709026}
{gr-qc/9709026};
K.S. Thorne 
\htmladdnormallink{gr-qc/9706057} 
{gr-qc/9706057}; 
E.E. Flanagan 
\htmladdnormallink{gr-qc/9706045}
{gr-qc/9706045}; 
A.G. Wiseman 
\htmladdnormallink{gr-qc/9704018}
{gr-qc/9704018}
; 
P.R. Brady and S.A. Hughes 
\htmladdnormallink{gr-qc/9704019}
{gr-qc/9704019};
D. Lai 
\htmladdnormallink{astro-ph/9605095}
{astro-ph/9605095}
.

[4] S.L. Shapiro 
\htmladdnormallink{gr-qc/9710094}
; G.B. Cook, S.L. Shapiro, and S.A. 
Teukolsky gr-qc/9512009.

[5] N. Andersson \htmladdnormallink{gr-qc/9706075}; 
J.L. Friedman and S.M. Morsink 
\htmladdnormallink{gr-qc/9706073}
{\http://xxx.lanl.gov/gr-qc/abs/9706073}; 
L. Lindblom, B.J. Owen, and S.M. Morsink 
\htmladdnormallink{gr-qc/9803053}
{\http://xxx.lanl.gov/gr-qc/abs/9803053}; 
S.M. Morsink, {\it Matters of Gravity} 
{\bf 10} 
\htmladdnormallink
{(Fall 1997)}
{http://vishnu.nirvana.phys.psu.edu/mog10/mog10.html}.

[6] S. Corley and T. Jacobson 
\htmladdnormallink{hep-th/9709166}
{\http://xxx.lanl.gov/gr-qc/abs//9709166}.

\vfill
\pagebreak
%%%%%%%%%%%%%%%%%%%%%%%%%%%%%%%%%%%%%%%%%%
\section*{\centerline {
Second International LISA Symposium}}
\addcontentsline{toc}{subsubsection}{\it  
Second International LISA Symposium, by Robin Stebbins}
\begin{center}
   Robin Stebbins, JILA, University of Colorado\\ 
\htmladdnormallink{stebbins@jila.colorado.edu}
{stebbins@jila.colorado.edu}
\end{center}
\parindent=0pt
\parskip=5pt

The Second International LISA Symposium was held at the
California Institute of Technology in Pasadena, 6-9 July 1998.
The symposium featured sessions on astrophysical sources of
gravitational waves detectable by an instrument like LISA,
relevant technology, data analysis and updates on other
detectors.  The oral sessions are summarized below.  Proceedings
of the symposium will be published by the American Institute of
Physics.  Bill Folkner, the scientific and local organizing
committees, and the supporting staff organized an action-packed
symposium.

Charles Elachi (JPL) started off the overview session by citing
the importance of LISA in the next decade and the importance of a
joint ESA-NASA mission.  Al Bunner (NASA) described selected and
candidate missions in the Structure and Evolution of the Universe
(SEU) program, and the decision-making process for selecting
future missions.  Rudiger Reinhard (ESA) described the status of
LISA within ESA's scientific program and the potential for ELITE,
a flight test of LISA technology.  Kip Thorne (Caltech) surveyed
the astrophysics and fundamental physics that will likely be
learned from LISA, and is unlikely to be learned from other
observations in the next 10 years.  Karsten Danzmann (Hannover)
compared and contrasted LISA and ground-based gravitational wave
detectors.  William Folkner (JPL) described the LISA mission
concept, and Robin Stebbins (JILA) summarized LISA's operation
and sensitivity.

In the first of two sessions on sources, the focus was massive
black holes (MBHs) at cosmological distances.  Roger Blandford
(Caltech), Doug Richstone (Michigan, IAS), Martin Haehnelt
(Cambridge) and Elihu Boldt (GSFC) described theoretical
scenarios and observational evidence leading to estimates of MBH
binary coalescence rates.  Steinn Sigurdsson (Cambridge) surveyed
the event rates for the inspiral of stellar remnants into MBHs.
The second session on sources included talks by Omer Blaes (U.C.
Santa Barbara), Sterl Phinney (Caltech), Ron Webbink (UIUC),
Dieter Hils (JILA), and Craig Hogan (Washington).  The topics
ranged from what we can learn from X-ray sources to discussions
of galactic binaries and possible primordial backgrounds.

There were three sessions on technology relevant to LISA.  The
first, with talks by David Robertson (Hannover), Paul McNamara
(Glasgow), Michael Peterseim (Hannover), Martin Caldwell (RAL),
Joe Giaime (JILA) and Joe Prestage (JPL), addressed topics in
interferometry, optics and timing.  In the second technology
session, Manuel Rodrigues (ONERA) and Stefano Vitale (Trento)
talked about inertial sensors.  Sasha Buchman (Stanford) and Mac
Keiser (Stanford) described relevant Gravity Probe B
technologies, and Dan DeBra (Stanford) surveyed space missions
which have used drag-free technology.  Salvo Marcuccio
(Centrospazio) and Michael Fehringer (Austrian Research Centre)
described different technologies for micronewton thrusters.  In
the last technology session, R. Turner (RAL) and Mike Sandford
(RAL) summarized structural, thermal and gravitational studies of
the LISA baseline design.  Yusuf Jafry (ESTEC), Mark Wiegand
(Bremen), David Robertson (Hannover) and Michael Peterseim
(Hannover) described the European LIsa TEchnology demonstration
satellite (ELITE) mission concept.

Curt Cutler (AEI) and Alberto Vecchio (AEI) led the data analysis
session by describing the angular resolution and astrophysical
parameter determination of the LISA and OMEGA missions.  Eric
Poisson (Guelph) described hierarchical search strategies to
identify the waveforms of stellar mass black holes spiraling into
massive black holes.  B. Sathyaprakash (Cardiff) examined a new
method to generate improved waveforms for MBH binary inspirals,
carry out an efficient search and extract source parameters.  A.
Sintes (AEI) described a scheme for removing coherent noise, and
Alessandra Papa (AEI) described a pattern recognition scheme for
identifying unexpected, but continuous gravitational wave
signals.  Finally, Massimo Tinto (JPL) presented a concept for
improving the detection sensitivity of a one arm interferometer
at selected Fourier components.

In a session on other gravitational wave detectors, the status of
the LIGO, GEO, TAMA, VIRGO and resonant mass gravitational wave
detectors was reported by David Shoemaker (MIT), Roland Schilling
(MPI Garching), K. Tsubono (Tokyo), Luca Gammaitoni (Perugia) and
E. Coccia (Rome), respectively.  John Armstrong (JPL) reported on
past and future sensitivity in spacecraft tracking experiments.
O. Aguiar (INPE, Brazil) reported on the resonant bar program in
Brazil.

\vfill
\pagebreak
%%%%%%%%%%%%%%%%%%%%%%%%%%%%%%%%%%%%%%%%%%
\section*{\centerline {
JILA Meeting on seismic isolation, test mass suspension,}\\
\centerline{and thermal noise issues for GW detectors}}
\addcontentsline{toc}{subsubsection}{\it  
JILA Meeting on seismic isolation et al., by Joe Giaime}
\begin{center}
   Joe Giaime, JILA, University of Colorado\\ 
\htmladdnormallink{giaime@jilau1.colorado.edu}
{giaime@jilau1.colorado.edu}
\end{center}
\parindent=0pt
\parskip=5pt

On August 11 - 12, 1998, about 35 people came to Boulder, Colorado to
attend the JILA Meeting on seismic isolation, test mass suspension,
and thermal noise issues for GW detectors.

The first session included talks on the measurements of ground noise
(Gabriela Gonzalez), seismic low-frequency ``pre-isolation" systems
(Ken Strain, David Shoemaker, Francesco Fidecaro), gravity noise (Kip
Thorne), and an interferometric GW detector experiment from 1970
(Judah Levine).  Next, there was a session on test mass suspension
systems (Mike Plissy, Norna Robertson), a novel method of thermal
noise monitoring and reduction (Yuri Levin), a proposal for cryogenic
thermal noise reduction (Warren Johnson), as well as a talk on the use
of a metal-insulator-metal diode as sensitive element for an
accelerometer (Alessandro Bertolini).  The topics for the third
session included passive and active seismic isolation in the GW band
for LIGO-I (Mark Barton), GEO 600 (Ken Strain), VIRGO (Giancarlo
Cella), and for the future (Joe Giaime, Dan Debra).

Immediately afterwards, on August 13 - 15, the third meeting of the
LIGO Science Collaboration was held.  About 80 collaborators were in
attendance.

Barry Barish gave an update on the state of the LIGO construction.
LIGO is 85\% complete as of August '98.  1999 will see the
interferometric detectors installed in the vacuum systems.  The system
will be commissioned in 2000, engineering tests will be conducted in
2001, and the initial coincidence data runs will begin in 2002.  He
also gave brief updates on the construction and system test progress
at the sites, and summarized the financial and staffing issues within
the LIGO Laboratory, and enumerated the membership of the LSC.
Raffaele Flaminio, Norna Robertson, and David McClelland presented
progress reports on VIRGO, GEO 600, and ACIGA.  VIRGO and GEO 600 have
start dates in the same approximate time frame as LIGO.

The three leaders of the experimental development groups presented
their reports.  These groups have been studying the technical issues
of future LIGO detector enhancements.  The focus was on preparing a
white paper for the PAC recommending a coherent research plan for
LIGO-related research that will lead to detector upgrades in 2004 and
2008 (or so.)  This goal has forced the three groups to consider
straw-man designs for the first upgrade and a fairly wide range of
possibilities for the second.  The first upgrade may include an
all fused silica double or triple pendulum similar to the one designed
for GEO600, a more powerful laser, some kind of low-frequency active
pre-isolation, and higher-quality optics.  More massive mirrors and
signal recycling are also being considered.  The second upgrade will
likely need even more challenging technology.  This may include
cryogenic suspensions and/or monitoring-balancing schemes to suppress
thermal noise, lower-frequency isolation, alternative test mass
materials with lower optical losses in order to tolerate still more
powerful lasers, and signal-tuned or adaptive interferometric length
detection schemes to maximize SNR for particular sources.  The three
presenters were David Shoemaker, Stochastic forces, Isolation Systems,
and Suspensions; Eric Gustafson, Sensing Noise - Lasers and Optics;
and Ken Strain, Interferometer Configurations.  Work is now underway
to produce an LSC-wide white paper.

The scientific goals driving the detector upgrade path were discussed
during two talks given by Peter Saulson and Kip Thorne.  It was
pointed out that for NS-NS binary coalescences, most of the
contribution to signal-to-noise ratio comes from the trough in the
noise curve between the falling pendulum thermal and radiation
pressure noise at low frequencies and the rising shot noise at high
frequencies, so a detector is optimized for this source by lowering
noise in the trough.  This is quantified in the form of a detection
range for a reasonable source SNR; the initial LIGO's NS-NS range
should be approximately to the VIRGO cluster, while the first upgrade
ought to improve that by a factor of 5 - 10.  Other sources can
involve different weighting, and various examples were discussed.

When Rainer Weiss introduced the content of the meeting, he
highlighted the need for the LSC experimenters to get more involved
with the three data- and astrophysics-oriented analysis groups,
especially the Detector Characterization group.  Reports from these
groups reflected quite a bit of progress over the last six months.
The Detector Characterization Group report, given by Daniel Sigg,
described the task's organization into data reduction, transient
analysis, performance characterization, and the production of
simulated data sets.  Bruce Allen gave the Astrophysical Signatures
group's talk, describing the tasks of using current astrophysical
knowledge to determine expected source characteristics, developing
algorithms, and developing the software implementation.  The
Validation and Detection Confidence group report, by Sam Finn,
explained this group's role as using the knowledge collected
by the previous two groups to determine answers to detection
confidence questions by statistical means.

Kathleen Johnson and Zeno Greenwood of the Center for Applied Physics
Studies at Louisiana Tech University and Alexander Sergeev of the
Institute of Applied Physics in Nizhny Novgorod, Russia, made
presentations to the LSC describing their institutions and proposing
to join the collaboration.  Later in the meeting,  during the business
meeting of the LSC Council both groups were approved for membership.
In addition, the Council appointed a nominating committee, which will
begin the process of replacing the current appointed leaders with
elected ones.

The conference viewgraphs are available in Acrobat form on the LIGO
web site,\\ 
\htmladdnormallink{\protect {\tt http://www.ligo.caltech.edu}}
{http://www.ligo.caltech.edu}.

\end{document}